# Temperature- and Polarization- Dependent Optical Properties of Single $Si_2Te_3$ Nanoplates

Jiyang Chen,[†] Romakanta Bhattarai,[†] Jingbiao Cui,[‡] Xiao Shen,[†] Thang Hoang[†,*]

[†] *Department of Physics and Materials Science, University of Memphis, TN 38152, USA*

[‡] *Department of Physics, The University of North Texas, TX76203, USA.*

[*]Email: tbhoang@memphis.edu

We report a combined experimental and computational study of the optical properties of individual silicon telluride ($Si_2Te_3$) nanoplates. The p-type semiconductor $Si_2Te_3$ has a unique layered crystal structure with hexagonal closed-packed Te sublattices and Si-Si dimers occupying octahedral intercalation sites. The orientation of the silicon dimers leads to unique optical and electronic properties. Two-dimensional $Si_2Te_3$ nanoplates with thicknesses of hundreds of nanometers and lateral sizes of tens of micrometers are synthesized by a chemical vapor deposition technique. At temperatures below 150 K, the $Si_2Te_3$ nanoplates exhibit a direct band structure with a band gap energy of 2.394 eV at 7 K and an estimated free exciton binding energy of 150 meV. Polarized reflection measurements at different temperatures show anisotropy in the absorption coefficient due to an anisotropic orientation of the silicon dimers, which is in excellent agreement with theoretical calculations of the dielectric functions. Polarized Raman measurements of single $Si_2Te_3$ nanoplates at different temperatures reveal various vibrational modes, which agree with density functional perturbation theory calculations. The unique structural and optical properties of nanostructured $Si_2Te_3$ hold great potential applications in optoelectronics and chemical sensing.

Semiconducting $Si_2Te_3$ nanostructures have recently emerged as materials that have potential applications as silicon-based devices because their unprecedented structural variabilities.[1-8] Previous studies of the electrical and optical properties of bulk $Si_2Te_3$ trace back to more than 50 years ago.[9-15] The p-type $Si_2Te_3$ semiconductor was reported to have both direct and indirect bandgap structures[9] with a hexagonal close-packed Te sublattice and Si atoms entangled as the Si-Si dimer between the Te layers.[10] The lattice constants of $Si_2Te_3$ were reported by Ploog et al.[11] as $a = 7.43$Å and $c = 13.48$Å. One- (1D) and two-dimensional (2D) $Si_2Te_3$ nanostructures were first synthesized by Kueleyan et al.[1] in 2015 by a chemical vapor deposition (CVD) technique. Shen et al.[5] also reported the variabilities of the electronic properties of nanostructured $Si_2Te_3$, including the tunable bandgap and band structures due to different Si-Si dimer orientations. Recent works by Wu et al.[2,16] and Wang et al.[8] investigated the growth mechanisms and optical properties of $Si_2Te_3$ nanowires and large $Si_2Te_3$ nanoplates (NPs). Our recent studies of $Si_2Te_3$ nanowires have also revealed interesting optical and electrical properties.[3,4] Electrical transport measurements on single $Si_2Te_3$ nanowires show a switching behavior that could have a potential application in memory devices.[4] The temperature, excitation power-dependent photoluminescence (PL), and decay dynamics of photoexcited carriers of $Si_2Te_3$ nanowires show very a long decay time of excitonic states and an indication of the temperature dependence of the Si-Si dimer orientation.[3] A theoretical calculation by Juneja et al.[17] shows a low thermal conductivity (2 W/mK at 1000 K) of n-doped $Si_2Te_3$, which results in a relatively high thermoelectric material figure of merit. Pressure-dependent studies have also shown interesting behaviors of the phase transition as well as bandgap modification.[6,7] Nevertheless, the study of the fundamental properties of nanostructured $Si_2Te_3$ is still in the early stage, especially the optical and electronic properties, for various reasons. For instance, one of the challenges in investigating the physical properties of nanostructured $Si_2Te_3$ is the stability of the material under ambient conditions. This is because the large surface-to-volume ratio at the nanoscale leads to a surface reaction with the water vapor in the atmosphere, resulting in a thin Te layer. Furthermore, the complication of the structural properties of $Si_2Te_3$ at low dimensions due to the orientation of the silicon dimers at different temperatures and strains could also lead to strikingly different optical or electronic properties.[5] On the other hand, these dependencies also offer an opportunity such that the properties of $Si_2Te_3$ are controllable using the set of parameters. Further, while $Si_2Te_3$ material can be processed using standard semiconductor techniques, it is highly sensitive to the environment, which offers an advantage for chemical sensing applications.[1,8] Indeed, $Si_2Te_3$ has one of the highest mechanical flexibilities among 2D materials ever reported.[6] Its interesting indirect-direct band gap transition with the application of mechanical uniaxial strain[6] and anisotropic optical dielectric properties[18,19] make this material even more interesting and provide promise for its potential applications.

In this study, we present the anisotropic optical properties of $Si_2Te_3$ NPs by combining a set of experiments and theoretical calculations. The reflection and Raman spectroscopies are explored by various measurement configurations. Temperature- and polarization-dependent measurements are also conducted. At temperatures below 150 K, a direct band structure is observed with a band gap value of 2.394 eV at 7 K. Polarized reflection measurements at different temperatures show an anisotropic behavior that is related to an anisotropy in the dielectric functions along different crystal directions due to the orientation of the Si-Si dimer. Similarly, polarized Raman scattering measurements also indicate an anisotropic absorption property. The measured polarized Raman and reflection results agree well with the first-principles calculations, which reveals the role of the Si-Si dimer orientations.

Si₂Te₃ NPs are synthesized by a CVD method. Specifically, Si$_2$Te$_3$ NPs are synthesized in a 2-inch tube furnace (MTI 1200X) by using tellurium (Te, 30 mesh, 99.997%, Sigma Aldrich) and silicon powders (Si, 325 mesh, 99%, Sigma Aldrich) as source materials. Te and Si powders are placed in a ceramic crucible and loaded into a high-temperature tube furnace (Fig. 1.a). The furnace is heated at a heating rate of 20°C/min. The precursors are placed inside the quartz tube and away from the heating area before the oven reaches the reaction temperature. When the temperature reaches the desired value, the precursors are pulled into the heating area for the reaction. A single-crystal silicon substrate is positioned downstream of the gas flow inside the oven and kept at 600°C while the source materials are heated to 850°C. Nitrogen is used as a carrier gas to keep the chamber pressure at 200 torr during the material growth. The nitrogen flow rate is set to 25 sccm by a mass flow controller. After the reaction is performed for a specific target time of 5 minutes, the furnace is quickly cooled to room temperature by opening the furnace's lid.

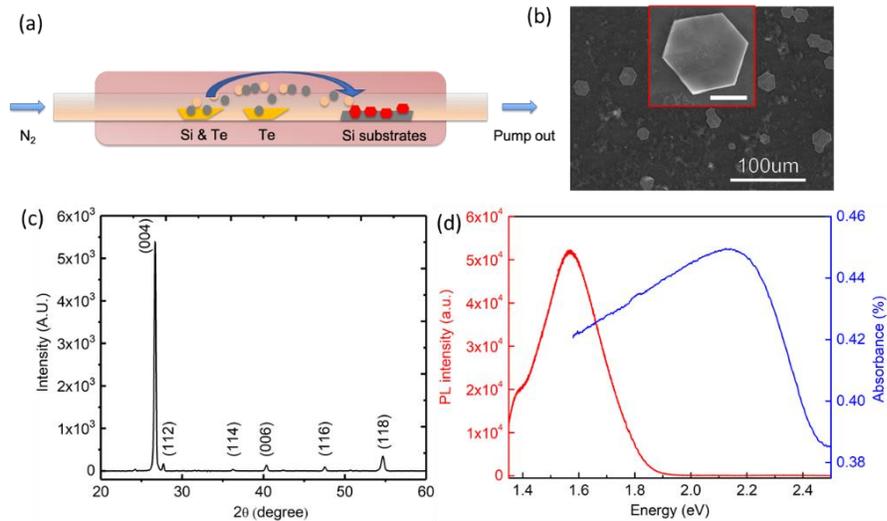

**Figure 1. Growth and characterization of Si$_2$Te$_3$ NPs.** (a) Schematic of the CVD synthesis process for synthesis of the Si$_2$Te$_3$ NPs. (b) SEM image of the Si$_2$Te$_3$ NPs, the inset (scale bar: 5 $\mu$m) is an enlarged image of a single NP. (c) Measured XRD pattern of the Si$_2$Te$_3$ NPs. (d) Typical absorption and emission curves of Si$_2$Te$_3$ NPs at room temperature.

In optical studies, the reflection and Raman measurements are conducted by using a spectrometer (Horiba iHR550) and charged coupled device camera (Horiba Jobin-Yvon Synapse). In the reflection measurement, the incident white light source is from a Xenon arc lamp with a broad, unpolarized spectrum from 350 to 1000 nm. The incident light is focused onto a single $Si_2Te_3$ NP via a 20x objective lens, and the reflection is collected by the same lens. For the polarized reflection measurement, the reflected light from a single $Si_2Te_3$ NP is analyzed by a linear polarizer (Thorlabs, LPVISC100-MP2) before being dispersed by the spectrometer. In the Raman measurements, a single $Si_2Te_3$ NP is excited by a 488 CW laser (Coherent Sapphire SF 488) via the 20x objective lens. The reflected Raman signal is collected by the same objective lens, dispersed, and analyzed by the spectrometer and camera. The 488 nm excitation laser is filtered by a notch filter (Semrock 488 nm laser clean-up filter) from the excitation side and by a longpass filter (488 nm ultrasteep longpass edge filter) from the detection side. For the polarized Raman measurements, the incident excitation laser is polarized by a half waveplate (Thorlabs AHWP05M-600) for different incident linear polarizations. For low-temperature measurements, the $Si_2Te_3$ NP temperature (7-293 K) is controlled by using a closed-cycle cryostat (Janis model CCS-XG-M/204N).

The morphology of the $Si_2Te_3$ NPs is characterized by scanning electron microscopy (SEM, Nova 650), as presented in Fig. 1.b. The NPs appear as a uniform hexagonal structure approximately 20 - 40 μm wide and 100 - 300 nm thick. Figure 1.c shows the X-ray diffraction pattern of the $Si_2Te_3$ NPs. The pattern indicates a series of diffraction peaks at 26.5°, 27.5°, 36.1°, 40.2°, 54.5°, and 60.5°, which correspond to the indices (004), (112), (114), (006), (008) and (118) of $Si_2Te_{3,\ respectively}$. The strong diffraction peak at 26.5° reveals that the growth of the $Si_2Te_3$ NPs is along the (001) direction. The results of the lattice structure study match well with previous studies.[1,16] Figure 1.d shows the absorption (blue) and the photoluminescence (PL) emission (red) curves of a single Si$_2$Te$_3$ NP. The Si$_2$Te$_3$ NP exhibits a strong and broad absorption spectrum with a maximum around 2.25 eV, which is close to the band gap of bulk Si$_2$Te$_3$ of 2.21 eV at room temperature.[2,9,12,20] The broad absorption spectrum that extends from orange to near-infrared is a result of the interband absorption of the Si$_2$Te$_3$ NP. The PL emission from the Si$_2$Te$_3$ NP also exhibits a broad spectrum in the near-infrared region, which indicates a defect-related emission. This result is similar to previous studies where the PL emission is mainly attributed to recombination at trap states above the valence band.[2,3,12,15]

Figure 2.a shows a typical low-temperature (7 K) unpolarized reflection spectrum of a single Si$_2$Te$_3$ plate. The value of the reflectance is calculated by the following equation: $Reflection = \frac{I - I_{noise}}{I_{sub} - I_{noise}}$. Here $I$, $I_{sub}$ and $I_{noise}$ are the reflected light intensities from the NP, silicon substrate, and camera thermal noise, respectively. To verify our measurement, Fig. 2.a also shows a reflection spectrum (dashed black curve) of the incident light on the silicon substrate ($I = I_{sub}$ in this case). As expected, the reflection from the silicon is close to 100%, after normalization to the same substrate. The reflection spectrum from a single Si$_2$Te$_3$ NP shows two interesting features. First, below the energy of approximately 2.38 eV the, reflectance value is smaller than 100%. This indicates that below this energy (2.38 eV), the incident photons are absorbed by the NP. This observation hints at the idea that the band gap of a Si$_2$Te$_3$ NP at 7 K is near 2.38 eV and that photons above this energy will be absorbed. The absorbed photons then promote electrons from the valence band to the conduction band or other defect centers to form excitonic states, which are subsequently annihilated to emit photons. Second, below the energy of 2.38 eV, the normalized reflectance value is greater than 100%, which indicates that additional light is emitted from the NP. It is very likely that the emitted light, such as PL, results from absorption at an energy larger

than the band gap and is followed by a re-emission process. For instance, a high-energy photon is absorbed to promote an electron to an excited state, which can then be relaxed or scattered by defects or phonons before returning to the ground state and emitting a photon at lower energy. This is consistent with previous observations that $Si_2Te_3$ has complicated defect/surface states that can result in PL emission in the 1.37 - 2.05 eV range, as shown in Fig. 1.d.[2,3,5]

To further understand the band structures of single $Si_2Te_3$ NPs, we performed the temperature-dependent reflection spectroscopy. Figure 2.b shows the reflection spectra from a single plate, which are measured at several temperatures between 7 K and 290 K. Indeed, the data in Fig. 2.b indicate several interesting properties. First, below 150 K, there are clear energy redshifts of the reflection spectra as the temperature increases. Specifically, when the temperature increases from 7 to 150 K, the crossing point between the reflection spectrum and the horizontal line (at 100% reflection) decreases from 2.38 eV to 2.32 eV. This is a strong indication of the bandgap shrinkage of $Si_2Te_3$ NPs because of the increasing temperature. Second, at temperature above 150 K, it is very abnormal that the crossing point between the reflection spectrum and the horizontal line (at 100% reflection) suddenly jumps to 2.44 eV and remains at this energy as the temperature approaches the room temperature of 290 K. This observation is consistent with previous studies[2,3] that the band structures of $Si_2Te_3$ nanostructures exhibit an abrupt change as the temperature transits through a critical temperature around 150 K. The mechanism of such a transition is not exactly clear at this moment. Nevertheless, there have been two possible explanations, including a modification of the band structures due to a reorientation of the Si-Si dimers[5] and the domination of the defect/surface-related states at high temperatures.[2,3]

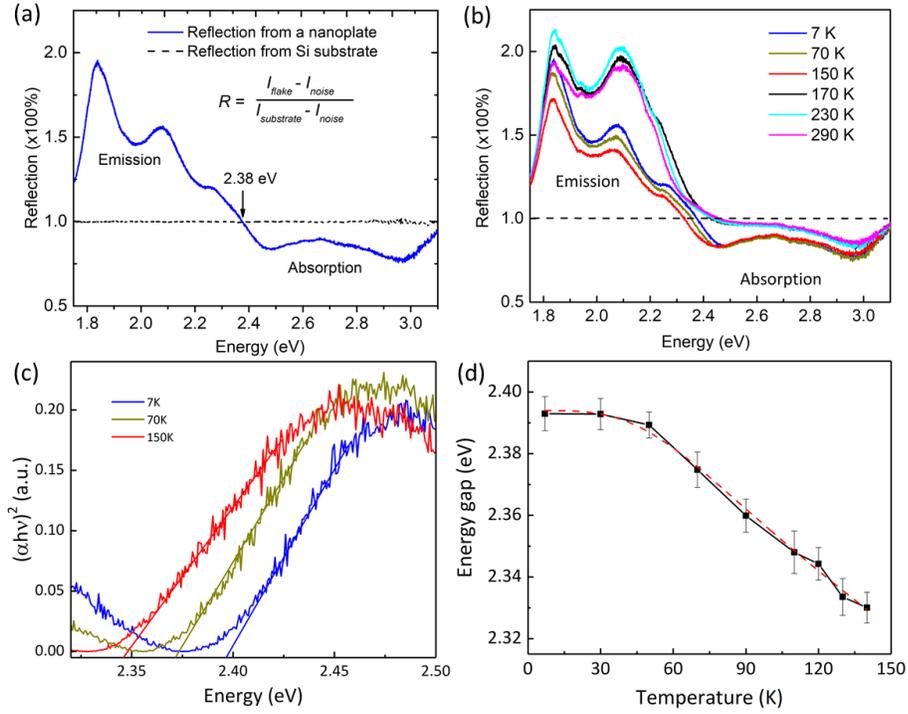

**Figure 2. Temperature-dependent reflection spectroscopy** (a) Reflection spectra from a single $Si_2Te_3$ NP at 7 K (solid, blue) and from a Si substrate (dashed, black). (b) Reflection spectra measured at different temperatures. The black dashed line indicates a reference value where the NP is absent. (c) Tauc plots at several representative temperatures. (d) Extracted band gap energy as a function of the temperature. The red dashed line is fit to the extended Varshni equation.

The reflection spectra at temperatures below 150 K presented in Fig. 2.a and 2.b hint at the idea of transition energies between the absorption to emission processes but do not yield accurate band gap energy values. To extract the value of the band gap energy, we employ a Tauc plot, which is often used to determine the optical band gaps of semiconductors.[21] In a Tauc plot, the absorbance is calculated from the reflectance, and the intercept with the x-axis of the linear fit in the absorption edge region yields the value of the band gap energy (Fig. 2.c). Further, in a Tauc plot the vertical axis represents the quantity $(\alpha h\nu)^n$, where $\alpha$ is the absorption coefficient and $h\nu$ is the photon energy. In this quantity, the exponent $n$ represents the nature of the transition, i.e $n = 2$ for direct band gap transitions and $n = 1/2$ for indirect band gap transitions. Previous work by Bletskan et al.[22] indicated an indirect band gap value of 2.13 eV for crystalline Si$_2$Te$_3$ at room temperature. Roma et al.[6] has also shown that the stress has a strong effect on the band gap energy as well as the nature of the transition (i.e., direct or indirect). Earlier work by Shen et al.[5] also demonstrated an interesting variability of the Si$_2$Te$_3$ band structures with respect to the orientation of the Si-Si dimers. In our measured data, while the Tauc plot for $n = 1/2$ (indirect transition) does not offer a meaningful result, the plot for $n = 2$ (direct transition) indicates a band gap value that is close to the crossing point between the reflection spectrum and the horizontal line, as described in Fig. 2.b. The fits to the Tauc plots in Fig. 2.c at temperatures of 7, 70 and 150 K result in band gap energy values of 2.394, 2.372 and 2.345 eV, respectively. It is interesting to note that in an earlier work by Wu et al.,[2] the authors showed a comparable redshift of approximately 30 meV for the free exciton energy level in a Si$_2$Te$_3$ NP ensemble when the temperature increases from 10 – 90 K. Indeed, if we compare the bandgap energy value of 2.394 eV at 7 K in this work with the free exciton energy level at 10 K as reported by Wu et al., one can estimate an approximate exciton binding energy of 150 meV for Si$_2$Te$_3$ NPs. Figure 2.d shows the band gap energy values obtained by fitting the Tauc plots for temperatures from 7-150 K. The red, dashed curve is the fit following the extended Varshni equation,[23] $E_g(T) = E_g(0) - \frac{\alpha T^4}{\beta+T^3}$, where $E_g(0)$ and $E_g(T)$ are the bandgap energies at temperatures 0 and T, respectively, and $\alpha$ and $\beta$ are constants to be determined. The fitting yields a band gap energy $E_g(0) = 2.396 \pm 0.002$ eV and $\alpha = 5.1730E - 4 \pm 1.8E - 5$, $\beta = 3.35E5 \pm 6.32E3$ for Si$_2$Te$_3$ NPs. It is also important to note that at temperatures above 150 K, the Tauc plots do not yield reliable fits to extract band gap energies. Similar to previous studies,[2,3] this result hints at an idea that at temperatures above 150 K the energy landscape of Si$_2$Te$_3$ NPs is complicated, and future additional measurements are needed to probe the band structures.

To obtain more information from the reflection from a single NP, Fig. 3 shows polarized reflection measurements. A single NP is excited by an unpolarized incident light source, and the reflected spectra are analyzed by a linear polarizer. Figure 3.a shows a clear analyzer angle dependence of the reflected intensity at 7 K. In this measurement, we are interested in the absorption part of the spectrum, i.e., at the energies above 2.39 eV. Figure 3.b plots the reflectance value at a fixed energy of 2.55 eV as a function of the linear analyzer angles at different temperatures. The highest reflectance is observed at 21º, and the lowest reflectance is observed at 111º, where the angles are relative to a predefined vertical direction. The result of the polarized reflection indicates an optical anisotropy behavior of the single $Si_2Te_3$ NP. The anisotropy of the optical absorption of a single $Si_2Te_3$ NP has a close relationship with the anisotropy of the dielectric functions along different crystal directions, as we will demonstrate later in a computational study. In addition, the anisotropy of the dielectric functions is closely related to the orientation of the Si-Si dimers; therefore, our observed absorption result aligns well with previous studies.[5,6,18]

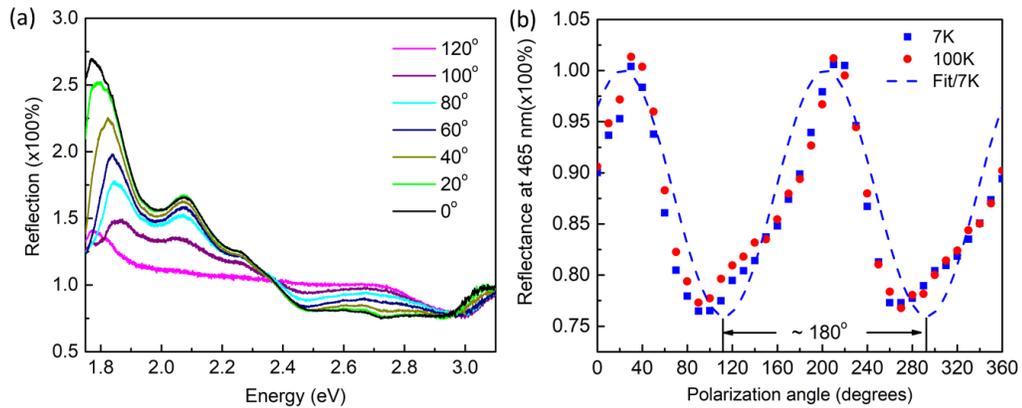

**Figure 3. Anisotropic optical absorption for single NPs.** (a) Reflection spectra at various analyzer angles at 7 K. (b) Reflectance measured at 2.55 eV at different temperatures and polarization angles. The dashed curve represents the fit by using a cosine function.

To further understand the anisotropy in optical processes, we perform temperature- and polarization-dependent Raman scattering measurements. Figure 4.a shows a typical Raman spectrum, which is measured at 20 K with a 30-second acquisition time. In this measurement, several resonant modes are observed. A doublet pattern appears at 130.5 and 154.1 cm$^{-1}$ from the $A_{1g}$ modes. This is similar to a previous measurement from single $Si_2Te_3$ NPs by Keuleyan et al.[1] In their work, Keuleyan et al. also report two additional peaks at 275.6 cm$^{-1}$ and 484 cm$^{-1}$ from $E_g$ modes. In our measurement, however, we observe several additional relatively weak peaks at 223.4, 336.5, 486.4, and 506.4 cm$^{-1}$. Figure 4.b presents a Raman scanning measurement from a single $Si_2Te_3$ NP at room temperature. The inset shows an optical image of the NP, and the dashed red line indicates the scanning across the NP. The scan step is 500 nm, which is approximately equal to the optical resolution, and the acquisition time is 10 seconds. We also note that to avoid heating and oxidation issues at room temperature, we avoid using a long acquisition time for each spot; therefore, several scattering peaks with lower intensity do not appear in this measurement. Nevertheless, the doublet pattern at 130.5 and 154.1 cm$^{-1}$ shows a persistently uniform scattering profile across the NP, indicating the high quality of the $Si_2Te_3$ sample. Further, we observe that the Raman frequencies at the edge of a single plate are essentially the same as those at the center of the NP. This observation indicates that, perhaps due to the relatively thick NPs used in our experiment, there is no strong evidence of edge states, as previously observed in other 2D nanomaterials such as $MoS_2$ and $WSe_2$.[24-28] In any case, if a single $Si_2Te_3$ NP is thinned down (by mechanical exfoliation, for example) to a thickness of a few monolayers, then it is possible to study the edge states in this materials, and indeed, this is the topic of our upcoming projects.

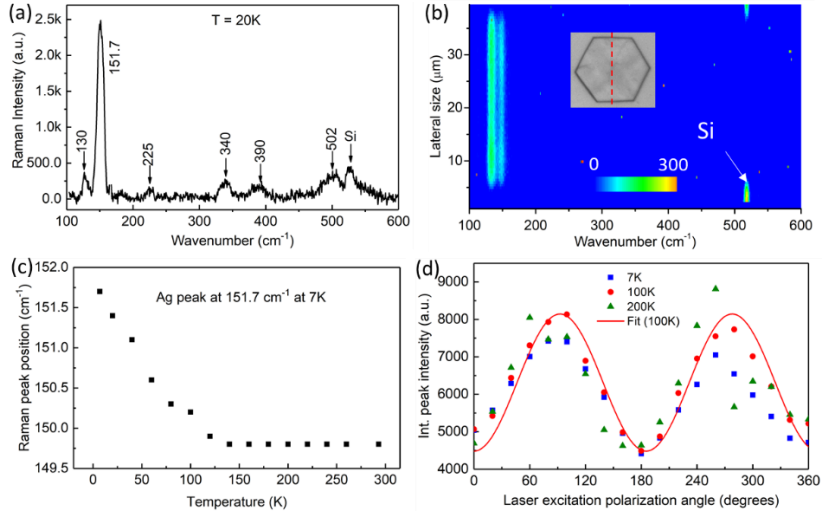

**Figure 4. Temperature-dependent and polarized Raman spectroscopy of single $Si_2Te_3$ NPs.** (a) Raman spectrum of a single $Si_2Te_3$ plate at 7 K. (b) Scanning Raman measurement across a single plate as indicated in the inset. (c) Temperature-dependent Raman peak position measured from 7 K to 290 K. (d) Polarization excitation-dependent Raman intensity. Both (c) and (d) correspond to the peak that appear at 151.7 cm$^{-1}$ at 7 K.

Figure 4.c shows the Raman peak position of an $A_{1g}$ peak from a single $Si_2Te_3$ NP at various temperatures. It is interesting that at temperatures from 7 to 150 K, the peak position shifts from 151.7 cm$^{-1}$ to 149.7 cm$^{-1}$ and remains at this frequency in the temperature range of 150-290 K. In addition, when the temperature increases, we observe that the shape of the Raman peak (not shown here) does not change. Indeed, Raman mode variation at different temperatures has been observed in other 2D materials and is related to thermal expansion.[29,30] However, for $Si_2Te_3$ NPs, there could be two reasons. First, the divergence behavior of the Raman peak at temperatures below 150 K is related to the thermal expansion of the relatively large thickness of the plate. As the temperature increases, the anharmonicity of the temperature expansion and the volume of the plate reach a saturation point before the lattice becomes stable. Second, the Raman shift is related to the band structure, which in turn is related to the orientation of the Si-Si dimers. The latter reason is supported by the reflection measurements and the band structure findings discussed above, where there is an abrupt change in the reflection and band gap energy at approximately 150 K. Further, the Raman frequency of a single Si$_2$Te$_3$ NP is closely related to the crystal structure, including the orientation of the Si-Si dimers, as we will discuss in the computational study below. In such a case, one would expect a polarization dependence of the Raman signal. Figure 5.d shows the polarization-excitation dependent Raman integrated intensity of the peak at 151.7 cm$^{-1}$ at several different temperatures. In this experiment, the laser excitation is linearly polarized by a half waveplate before being focused onto the sample. Similar to the reflection measurement data presented in Fig. 3.b, the Raman signal is polarized by 90º. The degree of polarization, which is defined as $P = \frac{I_{max}-I_{min}}{I_{max}+I_{min}}$, where $I_{max}$ and $I_{min}$ are the maximum and minimum intensities, respectively, is approximately 30%. The anisotropy in the Raman scattering intensity results from the anisotropic absorption coefficient of Si$_2$Te$_3$ NPs along different crystal directions, which is influenced by the orientation of the Si-Si dimers.

Finally, we will now discuss the computational study and relate our results with the experimental findings. We calculate the phonon spectrum using the first-principles density functional perturbation theory[31] as implemented in the Quantum Espresso package.[32] The Perdew-Burke-Ernzerhof (PBE)[33] exchange-correlation function of electrons under the generalized gradient approximation (GGA) method is used in the calculation. We use the norm-conserving (NC) pseudopotential[34] generated via the Rappe-Rabe-Kaxiras-Joannopoulos (RRKJ) method.[35] Plane-wave basis sets with 80 Rydberg (Ry) and 320 Ry cut-off energies are used for the expansion of the wave function and charge density, respectively. Raman intensity calculation is performed following a self-consistent calculation with a fully optimized structure. The convergence thresholds on the total energy and force for ionic minimization are set to 10$^{-4}$ Ry and 10$^{-3}$ Ry, respectively. Reciprocal space with a K-point grid of $5 \times 5 \times 3$ under the Monkhorst Pack scheme[36] is used for the integration over the Brillouin zone. Comparatively tight convergence criteria of $10^{-12}$ Ry and $10^{-14}$ Ry are set for self-consistent and phonon calculations, respectively. Finally, nonresonant Raman coefficients from the second-order response function are computed by taking the phonon wave vector[37] and applying the acoustic sum rule at the gamma point.

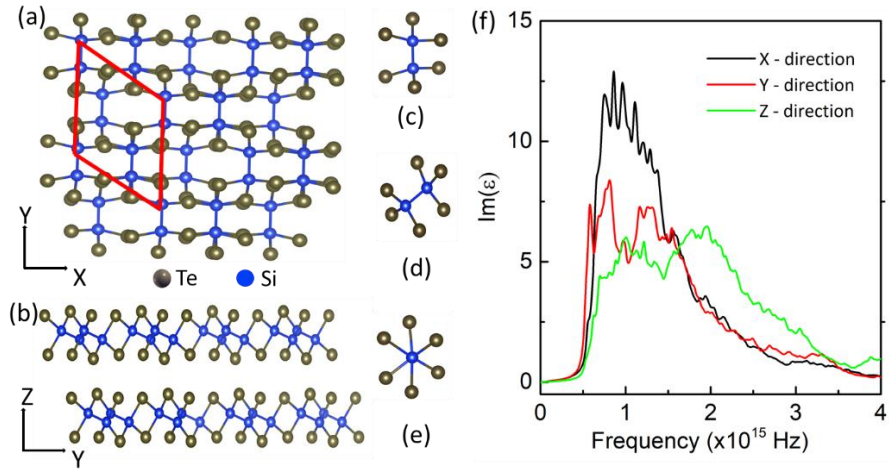

**Figure 5. Si$_2$Te$_3$ crystal structure and corresponding dielectric functions.** (a) Top view and (b) side view of the Si$_2$Te$_3$ crystal. (c-e) Different dimer orientations. Te and Si atoms are shown in gray and blue colors, respectively. (f) Calculated dielectric functions of Si$_2$Te$_3$ at different photon frequencies along three axes using the Bethe-Salpeter equation approach.[18]

Because the Si$_2$Te$_3$ NPs used in the experimental study are relatively large, in the simulation we consider a bulk Si$_2$Te$_3$ structure with lattice parameters $a = 7.50$ Å, $b = 8.63$ Å and $c = 13.97$ Å. The computational structure consists of 20 atoms with 12 Te and 8 Si atoms (Fig. 5.a-e). The y-axis is chosen to be parallel to the Si-Si dimers. The anisotropic optical properties of Si$_2$Te$_3$ are obtained by using the Bethe-Salpeter Equation (BSE) approach.[18] The dielectric responses of bulk Si$_2$Te$_3$ along all three axes with respect to the photon frequency are plotted in Fig. 5.f. Strong anisotropy is found, and it is mainly due to the anisotropic crystal structure along with the specific composition of the conduction and valence bands. The anisotropy in the dielectric responses along different crystal directions agrees well with the anisotropy in the absorption rates observed in the polarization-excitation dependent Raman measurement.

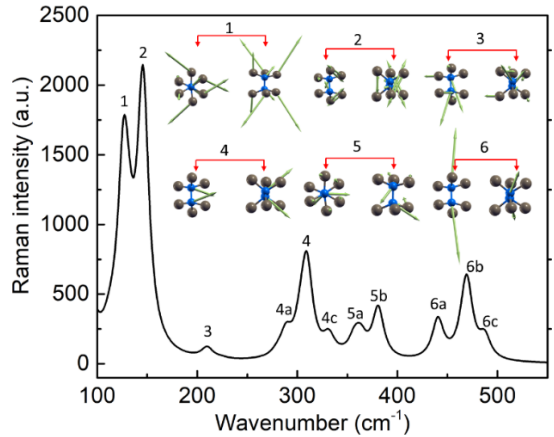

**Figure 6.** Calculated Raman spectrum of $Si_2Te_3$ showing the major and minor peaks along with the vibrational profiles of the respective major peaks (1-6). The insets show the corresponding orientations and vibrational directions.

The Raman intensities are computed for three different cases. First, all four Si-Si dimers are oriented horizontally along the same direction (in this case, along the y-axis), which we call the original structure with the minimum ground state energy. Second, one of the dimers is misaligned horizontally, and third, one of the dimers is misaligned vertically. The Raman frequency of a pure Si sample (521 cm$^{-1}$) is also calculated as a reference for our study. Our calculation shows that the horizontal misalignment of a dimer has no significant effect on the Raman peaks; however, the vertical misalignment results in one extra peak in the Raman spectrum (mode 1 in Fig. 6). After combining the data obtained for the three different cases mentioned above, we conclude that there are a total of six major peaks in the Raman spectrum at wavenumbers of 127 cm$^{-1}$, 144 cm$^{-1}$, 213 cm$^{-1}$, 310 cm$^{-1}$, 363 cm$^{-1}$, and 464 cm$^{-1}$, as depicted in Fig. 6. Our calculation result agrees well with the experimental Raman measurements, which indicate scattering peaks at 127 cm$^{-1}$, 152 cm$^{-}$

$^{-1}$, 225 cm$^{-1}$, 342 cm$^{-1}$, 398 cm$^{-1}$, and 499 cm$^{-1}$ (Fig. 4.a). Weaker shoulder peaks near the fourth, fifth, and sixth frequencies in the calculated data are not well resolved in the experimental spectrum, as shown in Fig. 4.a. We also analyze the vibrational profiles corresponding to these six major peaks in the insets of Fig. 6. The data show that the calculated Raman spectrum is strongly consistent with the experimental data, and the misalignment of the Si-Si dimers results in different vibrational modes.

In conclusion, in this work we study the anisotropic optical properties of single CVD-grown Si$_2$Te$_3$ NPs. Polarized reflection and Raman spectroscopies of single Si$_2$Te$_3$ NPs at various temperatures are presented. We observe the semiconducting Si$_2$Te$_3$ direct band gap at 2.394 eV at 7 K and an estimated free exciton energy of 150 meV. Polarized reflection and Raman measurement results are related to the anisotropy in the orientation of the Si-Si dimer and the dielectric response functions along different crystal directions. First-principles calculations of the dielectric functions and vibrational characteristics reveal excellent agreement with the experimentally observed data and provide a detailed picture of how the vibration and orientation of the Si-Si dimers determine the anisotropic properties of Si$_2$Te$_3$ NPs. Our study thus indicates interesting structural and optical properties of Si$_2$Te$_3$ nanostructures, which hold great potential for applications in optoelectronics and chemical sensing, especially because these materials have a clear advantage of compatibility with silicon-based devices.

## Acknowledgments

This work is supported by the National Science Foundation (NSF), DMR-1709528 (JC, JC and XS) and DMR-1709612 (TH), by the Ralph E. Powe Junior Faculty Enhancement Awards from Oak Ridge Associated Universities (to XS and TH). Computational resources were provided by the NSF XSEDE grant number TG-DMR 170064 and 170076 and the University of Memphis

High-Performance Computing Center (HPCC). TH acknowledges the FedEx Institute of Technology at the University of Memphis for its support.